\documentclass[a4paper,twoside,onecolumn,11pt]{book}
\usepackage[english]{babel}  
\usepackage{latexsym,mathptmx,multicol,ifthen}
\usepackage{longtable}

\usepackage{array}

\usepackage[scaled=0.92]{helvet}

\usepackage{amssymb}         
\usepackage{amsmath}

\usepackage{xr-hyper}

\makeatletter

\let\geq\geqslant  
\let\leq\leqslant  

\def\@removefromreset#1#2{\let\@tempb\@elt 
   \def\@tempa#1{@&#1}\expandafter\let\csname @*#1*\endcsname\@tempa
   \def\@elt##1{\expandafter\ifx\csname @*##1*\endcsname\@tempa\else
	 \noexpand\@elt{##1}\fi}%
   \expandafter\edef\csname cl@#2\endcsname{\csname cl@#2\endcsname}%
   \let\@elt\@tempb 
   \expandafter\let\csname @*#1*\endcsname\@undefined}

\@removefromreset{section}{chapter}

\@removefromreset{equation}{chapter}

\@removefromreset{subsection}{section}
\@removefromreset{subsubsection}{subsection}
\@removefromreset{paragraph}{subsubsection}
\@removefromreset{subparagraph}{paragraph}

\def\cstt{\@ifnextchar[{\cs@ktt}{\cs@tt}}

\def\cs@ktt[#1]#2{{\href{#1}{#2}}}  

\def\cs@tt#1{{\href{#1}{#1}}}  

\def\@csauth#1 #2/{{\def\csperrdimen{\kern 0.05 
     em}\textsc{\csperr{#1}} \textsc{\csperr{#2}}}\index{#2, #1}}

\def\@csauthcomma#1 #2/{{\def\csperrdimen{\kern 0.05 
     em}\textsc{\csperr{#1}} \textsc{\csperr{#2}}},\index{#2, #1}}	 
	 
     \def\@csasin[#1] #2/{
     \ifthenelse{\equal{#1}{1}}{\@csauth#2/}{}%
     \ifthenelse{\equal{#1}{2}}{\@csasintwo#2/}{}%
     \ifthenelse{\equal{#1}{3}}{\@csasinthree#2/}{}%
     \ifthenelse{\equal{#1}{4}}{\@csasinfour#2/}{}%
     \ifthenelse{\equal{#1}{5}}{\@csasinfive#2/}{}%
     \ifthenelse{\equal{#1}{6}}{\@csasinsix#2/}{}%
     \ifthenelse{\equal{#1}{7}}{\@csasinseven#2/}{}%
     \ifthenelse{\equal{#1}{8}}{\@csasineight#2/}{}%
     \ifthenelse{\equal{#1}{9}}{\@csasinnine#2/}{}%
     \ifthenelse{#1>9}{\typeout{Servono altri csasin nella riga nr. \the\inputlineno!}}{}%
     }	 

     \def\@csasintwo#1, #2/{%
     \@csauth#1/ \niceand \@csauth#2/,}

     \def\@csasinthree#1, #2, #3/{%
     \@csauth#1/, \@csauth#2/ \niceand \@csauth#3/,}

     \def\@csasinfour#1, #2, #3, #4/{%
     \@csauth#1/, \@csauth#2/, \@csauth#3/ \niceand \@csauth#4/,}

     \def\@csasinfive#1, #2, #3, #4, #5/{%
     \@csauth#1/, \@csauth#2/, \@csauth#3/,
     \@csauth#4/ \niceand \@csauth#5/,}

     \def\@csasinsix#1, #2, #3, #4, #5, #6/{%
     \@csauth#1/, \@csauth#2/, \@csauth#3/,
     \@csauth#4/, \@csauth#5/ \niceand  \@csauth#6/,}

     \def\@csasinseven#1, #2, #3, #4, #5, #6, #7/{%
     \@csauth#1/, \@csauth#2/, \@csauth#3/, \@csauth#4/,
     \@csauth#5/, \@csauth#6/ \niceand  \@csauth#7/,}

     \def\@csasineight#1, #2, #3, #4, #5, #6, #7, #8/{%
     \@csauth#1/, \@csauth#2/, \@csauth#3/, \@csauth#4/,
     \@csauth#5/, \@csauth#6/, \@csauth#7/ \niceand  \@csauth#8/,}

     \def\@csasinnine#1, #2, #3, #4, #5, #6, #7, #8, #9/{%
     \@csauth#1/, \@csauth#2/, \@csauth#3/, \@csauth#4/,
     \@csauth#5/, \@csauth#6/, \@csauth#7/,  \@csauth#8/
     \niceand  \@csauth#9/,}

     \def\niceand{\textsc{\& }}

     \def\isa[#1]{\bibitem{#1}}        

     \def\ti #1#2/{{\it\uppercase{#1}#2},}
     \def\tisim #1#2/{{\it\uppercase{#1}#2}}
     \def\bt #1/{{\it #1},}           
\let\np\noindent
\def\non{\nonumber\\ \relax}

\def\seepage#1{\relax}
\def\challengedif#1{ \relax}
\def\challengenor#1{ \relax}
\def\challengn{ \relax}

\def\iin#1{#1}
\def\iinn#1{#1}
\def\ii#1{\emph{#1}}
\def\se#1{}

\def\cp{{\hbox to 0pt{\;\;\;{.}\hss}}} 

\def\csd#1#2{\hbox{{${#1}$}}\:\hbox{{${\rm #2}$}}}

\let\cstimes\cdot
\def\csbul{\leavevmode\cssectioncolour{\rule[0.14em]{0.24em}{0.24em}}\kern
0.33em}
\let\e\varepsilon

\def\au{\@ifnextchar[{\@csau}{\@csau[]}}     

\def\auetal[#1] #2/{\bibitem{#1}\asietal#2/} 

\def\aualone[#1] #2/{\bibitem{#1}\asialone#2/}

\def\@csau[#1]#2/{\bibitem{#1}\asi#2/}
	 
\def\asietal#1 #2, #3/{{\def\csperrdimen{\kern 0.05 
     em}#1 \textsc{\csperr{#2}}} {\it\&\/} al.,\index{#2, #1}} 

\def\asialone#1/{{\def\csperrdimen{\kern 0.05 
     em}\textsc{\csperr{#1}}},\index{#1}} 

     \def\csperr#1{#1}  

\def\asi{\@ifnextchar[{\@csasin}{\@csauthcomma}} 

\def\isa[#1]{\bibitem{#1}}         

\def\ti #1#2/{{\it\uppercase{#1}#2},}
\def\tisim #1#2/{{\it\uppercase{#1}#2}}
\def\bt #1/{{\it #1},}           
\def\btsim #1/{{\it #1}}           
\def\jo #1/{#1}                  
\def\jotext #1/{{\it #1},}                     
\def\jotextsim #1/{{\it #1} }                  
\def\jotextend #1/{{\it #1}.}                  
\def\vo #1/{{{\bf #1},}}         
\def\pu #1/{#1,}                 
\def\puend #1/{#1.}              
\def\olo #1{{\oldstylenums{#1}}} 
\def\yrsim #1/{{\olo{#1}}}       
\def\yr #1/{{\olo{#1}},}         
\def\yrend #1/{{\olo{#1}}.}      
\def\pg #1/{p.~{\olo{#1}},}      
\def\pgsim #1/{p.~{\olo{#1}}}      
\def\pgend#1/{p.~{\olo{#1}}.}    

\def\pp #1-#2/{pp.~{\olo{#1}}\lower 0.2ex\hbox{--}{}{\olo{#2}},}
\def\ppend #1-#2/{pp.~{\olo{#1}}\lower 0.2ex\hbox{--}{}{\olo{#2}}.}

\def\yy #1-#2/{{\olo{#1}}\lower 0.2ex\hbox{--}{}{\olo{#2}}}
\def\yyend #1-#2/{{\olo{#1}}\lower 0.2ex\hbox{--}{}{\olo{#2}}.}

\skip\footins 30\p@ plus 4\p@ minus 2\p@

\def\normalparindent{\parindent 1.0em}   
\def\zeroparindent{\parindent \z@}

\parskip\z@                              

\@beginparpenalty -\@lowpenalty
\@endparpenalty   -\@lowpenalty
\@itempenalty     -\@lowpenalty

\def\cssectioncolour#1{#1}
\def\cssectioncolourh{\relax}

\def\section{%
  \par
  \penalty -50
  \vskip 0pt plus 40 pt
  \penalty 50
  \vskip 0pt plus -40 pt
  \vskip 0pt\relax
\@ifstar{\cs@s@section}{\cs@section}}

\def\cs@s@section{\@startsection{section}{1}{\z@}{-3.5ex plus -1ex minus 
 -.2ex}{2.3ex plus 
 .2ex}{\noindent\reset@font\sffamily\bfseries\cssectioncolourh}*}
 
\def\cs@section{\@startsection{section}{1}{\z@}{-3.5ex plus -1ex minus
   -.2ex}{2.3ex plus .2ex}{\noindent\reset@font\sffamily\bfseries\cssectioncolourh}}

   \def\subsection{%
     \par
     \penalty -50
     \vskip 0pt plus 40 pt
     \penalty 50
     \vskip 0pt plus -40 pt
     \vskip 0pt\relax
    \@startsection{subsection}{2}{\z@}{-3.25ex plus -1ex minus
     -.2ex}{1.5ex plus .2ex}{\reset@font\sffamily\bfseries\cssectioncolourh}}    
   \def\subsubsection{%
     \par
     \penalty -50
     \vskip 0pt plus 40 pt
     \penalty 50
     \vskip 0pt plus -40 pt
     \vskip 0pt\relax
    \@startsection{subsubsection}{3}{\z@}{-3.25ex plus -1ex 
	   minus -.2ex}{1.5ex 
	   plus.2ex}{\reset@font\small\sffamily\bfseries\itshape\cssectioncolourh}}

	   \def\thebibliography#1{
	      \list{{\reset@font\sffamily\bfseries\arabic{enumi}}}%
		    {\small\labelsep 0.5em\parindent 0em
		    \settowidth\labelwidth{#1}\leftmargin\labelwidth%
		    \frenchspacing\parindent=0pt\parskip=0pt
		    \advance\leftmargin\labelsep%
		    \itemindent \z@ \listparindent=0em
			    \parsep \z@%
		    \itemsep 0.3em plus \p@ minus \p@ 
		    \usecounter{enumi}}%
	   \def\endthebibliography{
	   \endlist\par} 
	      \sloppy
	      \sfcode`.=1000 \sfcode`,=1000 \sfcode`:=1000 \sfcode`;=1000
	      \sfcode`?=1000 \sfcode`!=1000 \sfcode`\.=1000}

	      \renewcommand{\@cite}[2]{\citefmp{Ref. #1}\if@tempswa{, #2}\else\fi}
	      
	      \def\citecs #1-#2/{{%
	      \def\@cite##1##2{{##1\if@tempswa , ##2\fi}}
	      \citefmp{Ref. \cite{#1}--\cite{#2}}%
	      \renewcommand{\@cite}[2]{\citefmp{Ref. ##1}\if@tempswa{, ##2}\else\fi}%
	      }}	      
	      
	      \def\citefmp#1{\mbox{}\marginpar[\raggedleft\hspace{0pt}\scriptsize\sf 
	      #1]{\raggedright\hspace{0pt}\scriptsize\sf #1}}

	      \def\fnsymbol#1{\expandafter\@fnsymbol\csname c@#1\endcsname}

	      \def\@fnsymbol#1{\ensuremath{\ifcase#1\or *\or **\or ***\or ****\or 
	      *****\or ******\or *******\else***\fi}}

      \def\@makefnmark{\hbox{$^{{}\:\@thefnmark}\m@th$}}                 

      \long\def\@makefntext#1{
      \def\strutdepth{\dp\strutbox} 
      \def\citefmp##1{\strut\vadjust{\kern-\strutdepth\vtop to 
      \strutdepth{\baselineskip\strutdepth\vss\llap{\scriptsize\sf##1\ \ \ \ \ 
      }\null}}}
      \def\fmp##1{\draftfinal{\strut\vadjust{\kern-\strutdepth\vtop to 
      \strutdepth{\baselineskip\strutdepth\vss\llap{\scriptsize\sf##1\ \ \ \ \ 
      }\null}}}{\typeout{C'e` ancora un fmp in una footnote, nella riga nr. \the\inputlineno}}}
      \def\bibitem##1{}
      \normalparindent
      \noindent
      \hbox{$\@thefnmark\;$}#1}                               

\makeatother

\usepackage{hyperref}

\begin{document} 
\bigskip
\bigskip
\bigskip
\bigskip
\bigskip
\bigskip
\bigskip
\bigskip
\bigskip
\bigskip

\section*{Maximum force and minimum distance: physics in limit statements}
\mark{{Maximum force and minimum distance}{Physics in limit statements}}

\medskip

\np Christoph Schiller

\np G \& D Research and Development

\np Prinzregentstra\ss e 159

\np 81677 M\"{u}nchen

\np Germany

\np christoph.schiller@motionmountain.net

\bigskip
\bigskip
\bigskip
\bigskip
\bigskip

\renewcommand{\baselinestretch}{1.0}\normalsize 
\subsubsection{Abstract}

\noindent The newly discovered principle of maximum force makes it possible to
summarize special relativity,\label{limiphi} quantum theory\se, and general
relativity in one fundamental limit principle each.  The three principles fully
contain the three theories and are fully equivalent to their standard
formulations.  In particular, using a result by Jacobson based on the Raychaudhuri equation, it is shown that a maximum force implies the field equations of general
relativity. The maximum force in nature is thus equivalent to the full 
theory of general relativity. Taken together, the three
fundamental principles imply a bound for every physical observable, from
acceleration to size.  The new, precise limit values differ from the usual
Planck values by numerical prefactors of order unity.  They are given here for
the first time.  Among others, a maximum force and thus a minimum length
imply that the non-continuity of space-time is an inevitable result of the
unification of quantum theory and relativity.

\bigskip
\bigskip
\bigskip

\subsubsection{Keywords} 

Maximum force, force limit, power limits, Planck limits, natural units

\bigskip
\bigskip
\bigskip
\bigskip

\newpage

\np Limit values for physical observables are often discussed in the
literature.\cite{ack} There have been studies of smallest distance, smallest
time intervals and smallest entropy values, as well as largest particle energy
and momentum values, largest acceleration values and largest space--time
curvature and other extreme values.\citecs me10-china/ Usually, these
arguments are based on limitations of measurement apparatuses tailored to
measure the specific observable under study.  In the following we argue that
all these limit statements can be deduced in a simpler way, namely 
by reformulating
the main theories of physics themselves as limit statements.  The limit value
for every physical observable then follows automatically, together with new,
corrected numerical prefactors.  This aim is achieved by 
condensing each domain of physics in a straightforward limit principle.  
Each principle is a limit statement that
limits the possibilities of motion in nature.  Before we discuss the last
principle that was missing in this chain, we summarize special relativity and
quantum theory in this way.  We then turn to general relativity, where we show
that it can be deduced from a new, equally simple principle.

\subsubsection{Special relativity in one statement}

It is well known that special relativity can be summarised by a single
statement on motion: \emph{There is a maximum speed in nature.} For all
systems,
\begin{equation}
    \quad v \leq c \quad.
\end{equation} 
A few well-known remarks set the framework for the later discussions.  The
speed $v$ is smaller than or equal to the speed of light for \emph{all}
physical systems;\footnote{A \ii{physical system} is a region of space--time
containing mass-energy, whose location can be followed over time and which
interacts incoherently with its environment.  With this definition, entangled
situations are excluded from the definition of system.} in particular, this
limit is valid both for composed systems as well as for elementary particles. 
The statement is valid for all observers.  No exception to the statement is
known.  Only a maximum speed ensures that cause and effect can be distinguished
in nature, or that sequences of observations can be defined.  The opposite
statement, implying the existence of (long-lived) tachyons, has been explored
and tested in great detail; it leads to numerous conflicts with observations.

The maximum speed forces us to use the concept of \emph{space-time} to describe
nature.  The existence of a maximal speed in nature also implies
observer-dependent time and 
space coordinates, to length contraction, time dilation\se, and all other
effects that characterise special relativity.  Only the existence of a maximum
speed leads to the principle of maximum aging that governs special relativity,
and thus at low speeds to the principle of least action.  In addition, only a
finite speed limit allows to define a \emph{unit} of speed.  If a speed limit
would not exist, no natural measurement standard for speed 
would exist in nature; in that case, speed would not be a measurable quantity.

Special relativity also limits the size of systems, independently of whether
they are composed or elementary.  Indeed, the speed limit implies that
acceleration $a$ and size $l$ cannot be increased independently without bounds,
as the two ends of a system must not interpenetrate.  The most important case
are massive systems, for which
\begin{eqnarray}
    && l \leq \frac{c^2}{a} \cp
    \label{eq:l2b}
\end{eqnarray} 
This size limit is also valid for the \emph{displacement} $d$ of a system, if
the acceleration measured by an external observer is used.  Finally, the limit implies
an `indeterminacy' relation:
\begin{eqnarray}
    &&\Delta l \; \Delta a \leq c^2
    \cp 
    \label{eq:l2bb}
\end{eqnarray} 
This is all textbook knowledge.

\subsubsection{Quantum theory in one statement}

In the same way, all of quantum theory can be summarised by a single statement
on motion: \emph{There is a minimum action in nature.} For all systems,
\begin{equation}
    \quad S \geq \frac{\hbar}{2} \quad.
\end{equation} 
Also this statement is valid both for composite and elementary systems.  The
action limit is used less frequently than the speed limit.  It starts from the
usual definition of the action, $S=\int (T-U) dt$, and states that between two
observations performed at times $t$ and $t + \Delta t$, even if the evolution
of a system is not known, the action is at least $\hbar/2$. The physical
action is a quantity that measures the change of state of a physical system.  In other words,
there is always a minimum change of state taking place between two observations
of a system.  The quantum of action expresses the well-known fundamental
fuzziness of nature at microscopic scale.

It is easily checked that no observation results in a smaller action value,
independently of whether photons, electrons\se, or macroscopic systems are
observed.  No exception to the statement is known.\footnote{In fact,
virtual particles can be seen as exceptions to this limit.}  
A minimum action has been
observed for fermions, bosons, laser beams, matter systems\se, and for any
combination of them.  The opposite statement, implying the existence of change
that is arbitrary small, has been explored in detail; Einstein's long
discussion with Bohr, for example, can be seen as a repeated attempt by
Einstein to find experiments which allow to measure arbitrary small changes in
nature.  In every case, Bohr found that this aim could not be achieved.

The minimum action value implies that in quantum theory, the three concepts of
state, measurement operation\se, and measurement result need to be
distinguished from each other; a so-called \emph{Hilbert space} needs to be
introduced.  The minimum action value can be used to deduce the uncertainty relation, the
tunnelling effect, entanglement, permutation symmetry, the appearance of
probabilities in quantum theory, the information theory aspect of quantum
theory and the existence of elementary particle reactions.  Details of this
discussion can be found in various textbooks.\cite{mec5a}

Obviously, the existence of a minimal or quantum of action was known right from
the beginning of quantum theory.  The quantum of action is at the basis of all
descriptions of quantum theory, including the many-path formulation and the
information-theoretic descriptions.  The existence of a minimum quantum of
action is completely equivalent to all standard developments.  In addition,
only a finite action limit allows to define a \emph{unit} of action.  If an
action limit would not exist, no natural measurement standard for action would
exist in nature; in that case, action would not be a measurable quantity.

The action bound $S\leq pd\leq mcd$, together with the quantum of action,
implies a limit on the displacement $d$ of a system between two observations:
\begin{equation}
     d \geq \frac{\hbar}{2 mc} \cp 
     \label{eq:l2a}
\end{equation} 
In other words, (half) the (reduced) Compton wavelength of quantum theory is
recovered as lower limit to the displacement of a system.  Since the quantum
\emph{displacement} limit applies in particular to an elementary system, the
limit is also valid for the \emph{size} of a composite system.  However, the
limit is \emph{not} valid for the size of \emph{elementary} particles.

The action limit of quantum theory also implies Heisenberg's well-known
\iin{indeterminacy relation} for the displacement and momentum of systems:
\begin{equation}
    \Delta d \; \Delta p \geq \frac{\hbar}{2}
    \cp
    \label{eq:l2aa}
\end{equation} 
It is valid both for massless and for massive systems.  All this is textbook
knowledge, of course.  One notes that by combining the limits (\ref{eq:l2b})
and (\ref{eq:l2a}) one obtains
\begin{eqnarray}
    &&a \leq \frac{2 mc^3}{\hbar} \cp
    \label{eq:l2ab}
\end{eqnarray} 
This \iinn{maximum acceleration} for systems in which gravity plays no role is
discussed in many publications.\cite{maxaccq} No experiment has ever reached
the limit, despite numerous attempts.

\subsubsection{General relativity in one statement}

Least known of all is the possibility to summarise general relativity in a
single statement on motion: \emph{There is a maximum force in
nature}. For all systems,
\begin{equation}
    \quad F \leq \frac{c^4}{4G} = \csd{3.0\cstimes 10^{43}}{N} \quad.
\end{equation} 
Let us explore the limit in some detail, as this formulation of general
relativity is not common.  (In fact, it seems that it has been discovered only
80 years after the general relativity has been around.\cite{me1} It might be
that the independent derivations of the present author and of Gary Gibbons have
been the first.)  The limit statement contains both the speed of light $c$ and
the constant of gravitation $G$; it thus indeed qualifies as a statement from
relativistic gravitation.  Like the previous limit statements, it is stated to
be valid for \emph{all} observers. The following discussion is given in more 
detail elsewhere.\cite{me20} 

The value of the maximum force is the mass--energy of a black hole divided by
its diameter.  It is also the surface gravity of a black hole times its mass. 
The force limit thus claims that no physical system of a given mass can
concentrated in a region of space--time \emph{smaller} than a (non-rotating)
black hole of that mass.  In fact, the mass--energy concentration limit can be
easily transformed by algebra into the force limit; both are equivalent.

It is easily checked that the maximum force is valid for all systems
\emph{observed} in nature, whether they are microscopic, macroscopic or
astrophysical.  Neither the `gravitational force' (as long as it is
operationally defined) nor the electromagnetic or the nuclear interactions are
found to ever exceed this limit.

The next aspect to check is whether a system can be \emph{imagined} that
exceeds the limit.  An extensive discussion shows that this is
impossible,\cite{me20} if the proper size of observers or test masses is taken
into account.  Even for a moving observer, when the force value is increased by
the (cube of the) relativistic dilation factor, or for an accelerating
observer, when the observed acceleration is increased by the acceleration of
the observer itself, the force limit must still hold.  However, no situations
allow to exceed the limit, as for high accelerations $a$, horizons appear at
distance $a/c^2$; since a mass $m$ has a minimum diameter given by $l\geq 4
Gm/c^2$, we are again limited by the maximum force.

The exploration of the force limit shows that it is achieved only on horizons;
the limit is reached in no other situation.  The force limit is valid for all
observers, all interactions\se, and all imaginable situations.

Alternatively to the maximum force limit, we can use as basic principle the
statement: \emph{There is a maximum power in nature.}
\begin{equation}
    P \leqslant \frac{c^5}{4G} = {9.1\cdot 10^{51}}{W}\quad.
\end{equation} 
The value of the force limit is the energy of a Schwarzschild black hole
divided by its diameter; here the `diameter' is defined as the circumference
divided by $\pi$.  The power limit is realized when such a black hole is
radiated away in the time that light takes to travel along a length
corresponding to the diameter.

In detail, both the force and the power limits state that the flow of momentum
or of energy \emph{through any physical surface} -- a term defined below -- of
any size, for any observer, in any coordinate system, never exceeds the limit
values.  Indeed, due to the lack of nearby black holes or horizons, neither
limit value is exceeded in any physical system found so far.  This is the case
at everyday length scales, in the microscopic world\se, and in astrophysical
systems.  In addition, even Gedanken experiments do not allow to exceed the
limits.\cite{me20} However, the limits become evident only when in such
Gedanken experiments the size of observers or of test masses is taken into
account.  Otherwise, apparent exceptions can be constructed; 
however, they are then
unphysical.\cite{me20}

\subsubsection{Deducing general relativity}

In order to elevate the force limit to a principle of nature,\label{mygrproof}
we have to show that in the same way that special relativity results from the
maximum speed, also general relativity results from the maximum force.

The maximum force and the maximum power are only realized on
horizons.\cite{stellartheory} (In fact, we can define the concept of horizon in
this way, and show that it is always a two-dimensional surface, and that it has
all properties usually associated with it.)  The type of horizon plays no role. 
Consider a (flat) horizon of area $A$ with surface gravity $a$ through which an
energy $E$ is flowing.  The force limit
\begin{equation}
    F\leqslant \frac{c^4}{4G}  
\end{equation}
in the case of a horizon implies the use of the equal sign.  For an energy $E$
flowing through a horizon surface $A$ we deduce
\begin{equation}
    \frac{E L}{A} = \frac{c^4}{4G} \quad, 
    \label{da191203}
\end{equation}
where $L$ is the proper length of the moving energy pulse or massive test body,
taken in the direction perpendicular to the horizon.  On a horizon, bodies feel
the surface gravity $a$; now, relativity shows that a pulse or body under
acceleration $a$ obeys $a L\leqslant c^{2}$.  Again, on a horizon, the extreme
case takes place, so that we have
\begin{equation}
    E = \frac{c^2}{4G} \, a\, A \quad,
    \label{db191203}
\end{equation}
where $a$ is the surface gravity of the horizon and $A$ its area.  The relation
can be rewritten for the differential case as
\begin{equation}
    \delta E = \frac{c^2}{4G} \, a\, \delta A   \quad.
    \label{dbb191203}
\end{equation}
In this way, the result can also be used for general horizons, 
such as horizons that are curved or time-dependent.%
\footnote{
Relation (\ref{dbb191203}) is well-known, though with
different names for the observables.  Since no communication is possible across
a horizon, also the detailed fate of energy flowing through a horizon is
unknown.  Energy whose detailed fate is unknown is often called \ii{heat}. 
Relation (\ref{dbb191203}) therefore states that the heat flowing through a
horizon is proportional to the horizon area.
When quantum theory is introduced into the discussion, the area of a horizon
can be called `entropy' and its surface gravity can be called `temperature';
relation (\ref{dbb191203}) can then be rewritten as
\begin{equation}
    \delta Q = T \delta S \quad.
\end{equation}  
However, this translation of the right hand side, which requires the quantum of
action, is unnecessary here.  We only cite it to show the relation to some of
the quantum gravity issues.} %

In a well known paper, Jacobson\cite{jacbis} has given a beautiful proof of a
simple connection: if energy flow is proportional to horizon area for all
observers and all horizons, then general relativity holds.  
To see the connection to general relativity, 
we generalize relation (\ref{dbb191203}) to
general coordinate systems and general energy-flow directions.
This is achieved by introducing tensor notation.

We start by introducing
the local boost Killing field $\chi$ that generates the horizon, with suitably
defined magnitude and direction; 
the surface gravity $a$ is the acceleration of the
Killing orbit with the maximal norm. 
This specifies what is meant by the term
`perpendicular' used above. We also introduce 
 the general surface element  $d \Sigma$. Using the 
 energy--momentum tensor $T_{ab}$
the left hand side of relation (\ref{dbb191203}) can then be rewritten
as
\begin{equation}
    \delta E = \int T_{ab} \chi^a d\Sigma^b   \quad.
\end{equation}	
Jacobson\cite{jacbis} then shows how the right hand side of 
relation (\ref{dbb191203})
can be rewritten, using the Raychaudhuri equation,
to give 
\begin{equation}
    a \delta A = \frac{c^2}{2 \pi} \int R_{ab} \chi^a d\Sigma^b   \quad,
\end{equation}
where $R_{ab}$ is the Ricci tensor.
The Ricci tensor describes how the shape of the horizon changes over space and
time.  

In summary, we get
\begin{equation}
    \int T_{ab}\chi^a d\Sigma^b = \frac{c^4}{8\pi G}\int
    R_{ab}\chi^a d\Sigma^b \quad,
    \label{dc191203}
\end{equation}
This equation and the local conservation of energy and momentum can both 
be satisfied only if
\begin{equation}
    T_{ab} = \frac{c^4}{8\pi G} \left (R_{ab}-(\frac{1}{2}R+\Lambda) g_{ab} 
    \right )
    \quad.
    \label{dd191203}
\end{equation}
These are the full field equations of general relativity, including the
cosmological constant $\Lambda$, which appears as an unspecified integration
constant.  By choosing a suitable observer, a horizon can be positioned at any
required point in space-time.  The equations of general relativity are thus
valid generally, for all times and positions.

In other words, \emph{the maximum force principle is a simple way to state that
on horizons, energy flow is proportional to area (and surface gravity)}.  This
connection allows to deduce the full theory of general relativity.  If no
maximum force would exist in nature, it would be possible to send any desired
amount of energy through a given surface, including any horizon.  In that case,
energy flow would not be proportional to area, black holes would not be of the
size they are\se, and general relativity would not hold.

The force and power bounds have important consequences.  In particular, they
imply statements on cosmic censorship, the Penrose inequality, the hoop
conjecture, the non-existence of plane gravitational waves, the lack of
space-time singularities, new experimental tests of the theory, and on the
elimination of competing theories of relativistic gravitation.  These
consequences are presented elsewhere.\cite{me20}

\subsubsection{Deducing universal gravitation}

Universal gravitation can be derived from the force limit in case that forces and
speeds are much smaller than the maximum values.  The first condition implies
$\sqrt{4GMa} \ll c^2$, the second $v \ll c$ and $al\ll c^2$.  To be concrete,
we study a satellite circling a central mass $M$ at distance $R$ with
acceleration $a$.  This system, with length $l=2R$, has only one characteristic
speed.  Whenever this speed $v$ is much smaller than $c$, $v^2$ must be
proportional both to $al=2aR$ and to $\sqrt{4GMa}$.  Together, this implies
$a=fGM/R^2$, where the numerical factor $f$ is not yet fixed.  
A quick check,\challengn
for example using the observed escape velocity values, shows that $f=1$.  Low
forces and low speeds thus imply that the inverse square law of gravity
describes the interaction between systems.  In other words, the force limit of
nature implies the universal law of gravity, as is expected.

\subsubsection{The size of physical systems in general relativity}

General relativity provides a limit on the \emph{size} of systems: there is a
limit to the amount of matter that can be concentrated into a small volume. 
The limit appears because a maximum force implies a limit to the depth of free
fall, as seen from an observer located far away.  Indeed, the speed of free
fall cannot reach the speed of light; the ever-increasing red-shift during
fall, together with spatial curvature, then gives an effective maximum depth of
fall.  A maximum depth of fall gives a minimum size or diameter $l$ of massive
systems.  To see this, we rewrite the force limit in nature as
\begin{equation}
    \frac{4Gm}{c^2} \leq \frac{c^2}{a} \cp
\end{equation} 
The right side is the \emph{upper} size limit of systems from special
relativity.  The left side is the Schwarzschild length of a massive system. 
The effects of space-time curvature make this length the \emph{lower} size
limit of a physical system:
\begin{equation}
    l \geq \frac{4Gm}{c^2}\cp
    \label{eq:l2c}
\end{equation} 
The size limit is only achieved for black holes, those well-known systems which
swallow everything that is thrown into them.  It is fully equivalent to the
force limit.  All \emph{composite} systems in nature comply with the lower size
limit.  Whether elementary particles fulfil or even achieve this limit remains
one of the open issues of modern physics.  At present, neither experiment nor
theory allow clear statements on their size.  More about this below.

General relativity also implies an `indeterminacy relation':\cite{grindet}
\begin{equation}
    \frac{\Delta E}{\Delta l} \leq
    \frac{c^4}{4G} \cp
    \label{eq:l2cc}
\end{equation} 
Since experimental data is available only for composite systems, we cannot say
yet whether this inequality also holds for elementary particles.  The relation
is not as popular as the previous two, but common knowledge in general
relativity.  In fact, testing the relation, for example with binary pulsars,
might lead to new tests to distinguish general relativity from competing
theories.

\subsubsection{Deducing limit values for all physical observables}

The maximum force of nature is equivalent with general relativity and includes
universal gravity.  As a result, three simple statements on nature can be made:
\begin{eqnarray}
    &\hbox{quantum theory:} &\quad S \ \geq \frac{\hbar}{2} \non
    &\hbox{special relativity:}&\quad v \ \leq c \non
    &\hbox{general relativity:}&\quad F \ \leq  \frac{c^4}{4G} 
    \label{eq:l1}
\end{eqnarray} 
The limits are valid for all physical systems, whether composed or elementary,
and are valid for all observers.  One notes that the limit quantities of
special relativity, quantum theory\se, and general relativity can also be seen
as the right hand side of the respective indeterminacy relations.  Indeed, the
set (\ref{eq:l2bb}, \ref{eq:l2aa}, \ref{eq:l2cc}) of indeterminacy relations or
the set (\ref{eq:l2b}, \ref{eq:l2a}, \ref{eq:l2c}) of length limits are fully
equivalent to the three limit statements (\ref{eq:l1}).\challengn Each set of
limits can be seen as a summary of a section of twentieth century physics.

If the three fundamental limits are combined, a limit for a number of
observables for physical systems appear.  The following limits are valid
generally, both for composite and for elementary systems:
\begin{alignat}{3}
    &\hbox{time interval:} &\quad t &\geq \sqrt{\frac{2G\hbar}{ c^5 }} 
    &=\;\;\;\;& \csd{7.6 \cstimes 10^{-44}}{s}  \displaybreak[0]\\
    &\hbox{time distance product:}& \quad td &\geq{\frac{2G\hbar}{ c^4}}
     &=\;\;\;\;& \csd{1.7 \cstimes 10^{-78}}{sm}    \displaybreak[0]\\
    &\hbox{acceleration:} &\quad a &\leq \sqrt{\frac{c^7 }{2 G\hbar }} 
    &=\;\;\;\;& \csd{4.0 \cstimes 10^{-51}}{m/s^2}   \displaybreak[0]\\
    &\hbox{power, luminosity:} & \quad P &\leq \frac{c^5}{4G} &=\;\;\;\;&
    \csd{9.1 \cstimes 10^{51}}{W} \displaybreak[0]\\
    &\hbox{angular frequency:} & \quad \omega &\leq 2 \pi\sqrt{\frac{c^5
    }{2G\hbar}}&=\;\;\;\;& \csd{8.2 \cstimes 10^{43}}{/s} \displaybreak[0]\\
    &\hbox{angular momentum:} & \quad D &\geq \frac{\hbar}{2} &=\;\;\;\;& 
    \csd{0.53
    \cstimes 10^{-34}}{Js}  \displaybreak[0]\\
    &\hbox{entropy:} & \quad S &\geq k
    &=\;\;\;\;&Ê\csd{13.8}{yJ/K} \displaybreak[0]\\
    \intertext{With the additional knowledge that in nature, space
    and time can mix, one gets} 
    &\hbox{distance:} & \quad d &\geq \sqrt{\frac{2G\hbar}{c^3 }} &=\;\;\;\;&
    \csd{2.3 \cstimes 10^{-35}}{m} \displaybreak[0]\\
    &\hbox{area:} & \quad A &\geq {\frac{2G\hbar }{c^3 }}  &=\;\;\;\;& \csd{5.2
    \cstimes 10^{-70}}{m^2} \displaybreak[0]\\
    &\hbox{volume} & \quad V &\geq \left ({\frac{2G\hbar }{c^3 }} \right)^{3/2}
    &=\;\;\;\;& \csd{1.2  \cstimes 10^{-104}}{m^3} \displaybreak[0]\\
    &\hbox{curvature:} & \quad K &\leq \frac{c^3 }{2G\hbar } &=\;\;\;\;& \csd{1.9
    \cstimes 10^{69}}{/m^2} \displaybreak[0]\\
    &\hbox{mass density:} & \quad \varrho &\leq \frac{c^5 }{8 G^2 \hbar }
    &=\;\;\;\;& \csd{6.5 \cstimes 10^{95}}{kg/m^3} 
\end{alignat} 
\np Of course, speed, action\se, and force are limited as already stated. 
Within a small numerical factor, for every physical observable these limits
correspond to the Planck value.  (The limit values are deduced from the
commonly used Planck values simply by substituting $G$ with $4G$ and $\hbar$
with $\hbar/2$.)  These values are the true \ii{natural units} of nature.  In
fact, the most aesthetically pleasing solution is to redefine the usual Planck
values for every observable to these extremal values by absorbing the numerical
factors into the respective definitions.\index{natural units}\index{units,
natural} In the following, we call the redefined limits the \ii{(corrected)
Planck limits} and assume that the factors have been properly included.  In
other words, \emph{the natural unit or (corrected) Planck unit is at the same
time the limit value of the corresponding physical observable.}

Most of these limit statements are found scattered around the literature,
though the numerical prefactors are often different.  The existence of a
smallest measurable distance and time interval of the\cite{me10} order of the
\iin{Planck values} are discussed in quantum gravity and string theory.  A
largest curvature has been discussed\cite{maxcurv} in quantum gravity.  The
maximal mass density appears in the discussions on the energy of the vacuum.

With the present deduction of the limits, two results are achieved.  First of
all, the various arguments found in the literature are reduced to three
generally accepted principles.  Second, the confusion about the numerical
factors is solved.  During the history of Planck units, the numerical
factors have greatly varied.  For example, Planck did not include the
factors of $2\pi$.  The fathers of quantum theory forgot the 1/2 in the
definition of the quantum of action.  And the specialists of relativity did not
underline the factor 4 too often.  With the present framework, the issue of the
correct factors in the Planck units can be considered as settled.

The three limits of nature (\ref{eq:l1}) result in a minimum distance and a
minimum time interval.  These minimum intervals result from the unification of
quantum theory and relativity.  They do not appear if the theories are kept
separate.  In short, unification implies that there is a smallest length in
nature.  The result is important: \emph{the formulation of physics as a set of
limit statements shows that the continuum description of space and time is not
correct.} Continuity and manifolds are only approximations valid for large
action values, low speed and low force values.  However, the way that a minimum
distance leads to a homogeneous and isotropic vacuum is still an open issue at
this point.

\subsubsection{Mass and energy limits}

Mass plays a special role in all these arguments.  The set of limits
(\ref{eq:l1}) does not allow to extract a limit statement on the mass of
physical systems.  To find one, the aim has to be restricted. 

The Planck limits mentioned so far apply for \emph{all} physical systems,
whether they are composed or elementary.  Additional limits can only be found
for elementary systems.  In quantum theory, the distance limit is a size limit
only for \emph{composed} systems.  A particle is elementary if the system size
$l$ is smaller than any conceivable dimension:
\begin{eqnarray}
    &&\hbox{for elementary particles:}\quad l \leq \frac{\hbar}{2 mc} \cp
\end{eqnarray} 
By using this new limit, valid only for elementary particles,  the
well-known mass, energy\se, and momentum limits are found:
\begin{eqnarray}
    &&\hbox{for elementary particles:}\quad m \leq \sqrt{\frac{\hbar c }{ 8G}}
    \quad =  \csd{7.7 \cstimes 10^{-9}}{kg} = \csd{0.42
    \cstimes 10^{19}}{GeV/c^{2}} \non
    &&\hbox{for elementary particles:}\quad E \leq \sqrt{\frac{\hbar c^5 }{ 8G}}
    \quad = \csd{6.9 \cstimes 10^{8}}{J} = \csd{0.42
    \cstimes 10^{19}}{GeV} \non
    &&\hbox{for elementary particles:}\quad p \leq \sqrt{\frac{\hbar c^3 }{ 8G}}
    \quad = \csd{2.3}{kg\,m/s}= \csd{0.42
    \cstimes 10^{19}}{GeV/c} 
\end{eqnarray} 
These single particle limits, corresponding to the corrected Planck mass,
energy\se, and momentum, were already discussed in 1968 by \iinn{Andrei
Sakharov}, though again with different numerical prefactors.  They
are\cite{asakh} regularly cited in elementary particle theory.  Obviously, all
known measurements comply with the limits.

\subsubsection{Corrected electromagnetic Planck limits}

The discussion of limits can be extended to include electromagnetism.  Using
the (low-energy) electromagnetic coupling constant $\alpha$, one gets the
following limits for physical systems interacting electromagnetically:
\begin{alignat}{2}
    &\hbox{electric charge:} & \quad q &\geq \sqrt{\strut 4\pi\e_{\rm o}\alpha c\hbar} = 
    e = \csd{0.16}{aC}   \displaybreak[0]\\
    &\hbox{electric field} &\quad E &\leq \sqrt{\frac{c^7}{64\pi\e_{\rm
    o}\alpha \hbar G^2}} = \frac{ c^4}{4Ge}=\csd{2.4\cstimes10^{61}}{V/m} \displaybreak[0]\\
    &\hbox{magnetic field (flux density):} & \quad B &\leq
    \sqrt{\frac{c^5}{64\pi\e_{\rm o}\alpha\hbar G^2}} = \frac{
    c^3}{4Ge}=\csd{7.9\cstimes10^{52}}{T} \displaybreak[0]\\
    &\hbox{voltage:} & \quad U &\leq \sqrt{\frac{c^4}{32\pi \e_{\rm o}\alpha
    G}} = \frac{1}{e} \sqrt{\frac{\hbar c^5}{8G}} = \csd{1.5\cstimes 10^{27}}{V}
    \displaybreak[0]\\
    &\hbox{inductance:} &\quad L &\geq \frac{1}{8\pi\e_{\rm 
    o}\alpha}\sqrt{\frac{
    2\hbar G}{c^7}} = \frac{1}{e^2}\sqrt{\frac{\hbar^3 G}{2 c^5}} =
    \csd{4.4\cstimes10^{-40}}{H} \displaybreak[0]\\
    \intertext{With the additional assumption that in nature at most one
    particle can occupy one Planck volume, one gets}
    &\hbox{charge density:} & \quad \varrho_{\rm e} &\leq \sqrt{\frac{\pi \e_{\rm
    o}\alpha}{2 G^3}} \; \frac{c^5}{\hbar} = e\sqrt{\frac{c^9}{8 G^3\hbar^3}}
    =\csd{1.3\cstimes 10^{85}}{C/m^3} \displaybreak[0]\\
    &\hbox{capacitance:} &\quad C &\geq 8\pi\e_{\rm o}\alpha\sqrt{\frac{2\hbar
    G}{c^3}} = e^2\sqrt{\frac{8 G}{c^5 \hbar}} =\csd{1.0\cstimes10^{-46}}{F}
    \intertext{For the case of a single conduction channel, one gets}
    &\hbox{electric resistance:} & \quad R &\geq \frac{1}{8\pi\e_{\rm o}\alpha
    c} = \frac{\hbar}{2 e^2} = \csd{2.1}{k\Omega} \displaybreak[0]\\
    &\hbox{electric conductivity:} & \quad G &\leq   8\pi\e_{\rm o}\alpha c =
    \frac{2 e^2}{\hbar} = 
    \csd{0.49}{mS} \displaybreak[0]\\
    %
    &\hbox{electric current:} & \quad I &\leq \sqrt{\frac{2\pi\e_{\rm o}\alpha
    {c^6}}{G}} = e \sqrt{\frac{c^5}{2 \hbar G}}=\csd{7.4\cstimes10^{23}}{A}
\end{alignat} 
Several electromagnetic limits, such as the magnetic field limit, play a role
in the discussion of extreme stars and black holes. 
The maximal electric field plays a role in the theory of gamma ray
bursters.\cite{ruff} Also the restriction of limit values for current,
conductivity\se, and resistance to single channels is well known in the
literature.  Their values and effects have been studied extensively in the
1980s and 90s.\cite{beni}

The observation of collective excitations in semiconductors with charge $e/3$,
and of quarks does not invalidate the charge limit for physical systems.  In
both cases there is no physical system -- defined in the sense give above as
localized mass-energy interacting incoherently with the environment -- with
charge $e/3$. 

\subsubsection{Thermodynamics}

Also thermodynamics can be summarized in a single statement on motion:
\emph{There is a smallest entropy in nature.}
\begin{equation}
	S \geq k \quad.
\end{equation} 
The result is almost 100 years old; it was stated most clearly by \iinn{Leo
Szilard}.\cite{szila} In the same way as in the other fields of physics, also
this result can be phrased as a indeterminacy relation:
\begin{eqnarray}
    &&\Delta\frac{1}{T} \; \Delta U
    \geq k \cp
    \label{eq:lther}
\end{eqnarray} 
This relation has been already given by \iin{Bohr} and was discussed by
\iin{Heisenberg} and many others.\cite{thermocic} It is mentioned here in order
to complete the list of indeterminacy relations and fundamental constants.
With the single particle limits, the entropy limit leads to an upper limit
for temperature:
\begin{eqnarray}
    &&T \leq \sqrt{\frac{\hbar c^5}{8 G k^2}}
    = \csd{5.0\cstimes 10^{31}}{K} \cp
\end{eqnarray} 
It corresponds to that temperature where the energy of every elementary
particle is given by the (corrected) Planck energy.

\subsubsection{Paradoxes and curiosities about Planck limits}

The (corrected) Planck limits are statements about properties of nature.  There
is no way to measure values exceeding these limits, whatever experiment is
performed.  As can be expected, such a claim provokes the search for
counter-examples and leads to many paradoxes. 

\csbul The minimal angular momentum might surprise at first, especially when
one thinks about spin zero particles.  However, the angular momentum of the
statement is \emph{total} angular momentum, including the orbital part with
respect to the observer.  The total angular momentum is never smaller than
$\hbar/2$.

\csbul If any interaction is stronger than gravity, how can the maximum force
be determined by gravity alone, which is the weakest interaction?  It turns out
that in situations near the maximum force, the other interactions are
negligible.  This is the reason that gravity must be included in a unified
description of nature. 

\csbul On first sight, it seems that electric charge can be used in such a way
that the acceleration of a charged body towards a charged black hole is
increased to a value exceeding the force limit.  However, the changes in the
horizon for charged black holes prevent this.

\csbul Some limits are of interest if applied to the \iin{universe} as a whole,
such as the luminosity limit (which, together with the age and size of the
universe, explains why the sky is dark at night) and the curvature limit (which
is of importance near the big bang).  The angular rotation limit also provides
a limit on the rotation of the observed matter in the sky.  

\csbul The limit on power can be challenged when several sources almost as
bright as the maximum are combined into one system.  Two cases need to be
distinguished.  If sources are far away, the luminosities cannot added up
because the energy from the sources arrives at different times; for bright
sources which are near each other, the combination forms a black hole or
at least prevents all radiation to be emitted by swallowing some of it among the
sources.

\csbul A precise definition of power, or energy per time, implies the
definition of a surface for which the power is measured.  This surface must be
physical, i.e.~must not cross horizons nor have curvatures larger than the
maximum possible value.  Otherwise, counter-examples to the power limit can be
constructed.

\csbul The general connection that to every limit value in nature there is a
corresponding indeterminacy relation is also valid for electricity.  Indeed,
there is an indeterminacy relation for capacitors of the form
\begin{eqnarray}
    \Delta C \; \Delta U \geq e
\end{eqnarray} 
where $e$ is the positron charge, $C$ capacity\se, and $U$ potential
difference, and one between electric current $I$ and time $t$
\begin{eqnarray}
    \Delta I \; \Delta t \geq e 
\end{eqnarray} 
and both relations are found in the literature.\cite{china}

\csbul The minimum area is twice the uncorrected Planck area.  This means that
the correct entropy relation for black holes should be $S/S_{\rm min}=A/2A_{\rm
min}$.  The factor 2 replaces the factor 4 that appears when the standard,
uncorrected Planck area is used.

\subsubsection{Some consequences of the limits}

The existence of limit values for the length observable (and all others) has
numerous consequences discussed in detail elsewhere.\cite{me10,me11bis} A few
are summarized here.

The existence of a smallest length -- and a corresponding shortest time
interval -- implies that no surface is physical if any part of it requires a
localization in space-time to dimensions smaller that the minimum length.  (In
addition, a physical surface must not cross any horizon.)  Only through this
condition unphysical counter-examples to the force and power limits mentioned
above are eliminated.  For example, this condition has been overlooked in
Bousso's older discussion of the Bekenstein's entropy bound\cite{boussobis} --
though not in his newer ones.

Obviously, a minimum length implies that \emph{space, time\se, and space-time
are not continuous.} The reformulation of general relativity and quantum theory
with limit statements makes this result especially clear.  The result is thus a
direct consequence of the unification of quantum theory and general relativity.
No other assumptions are needed.

The corrected value of the Planck length should also be the expression that
appears in the so-called theories of `doubly special relativity'.\cite{dsrbis}
They try to expand special relativity in such a way that an invariant length
appears in the theory.

A force limit in nature implies that no physical system can be smaller than a
Schwarzschild black hole of the same mass.  The force limit thus implies that
\emph{point particles do not exist.} So far, this prediction is not
contradicted by observations, as the predicted sizes are so small to be outside
experimental reach.  If quantum theory is taken into account, this bound is
sharpened.  Due to the minimum length, elementary particles are now predicted
to be larger than the corrected Planck length.  Detecting the sizes of
elementary particles would allow to check the force limit; this might be
possible with future electric dipole measurements.\cite{me10}

In addition, a limit to observables implies that at Planck scales no physical
observable can be described by real numbers, that no low-energy symmetry is
valid\se, and that matter and vacuum cannot be
distinguished.\cite{me10,me11bis} 

\subsubsection{Minimum length and measurements}

\noindent A smallest length, together with the limits for all other observables,
has a central consequence.  \emph{All measurements are limited in precision.} A
precise discussion\cite{me10} shows that measurement errors increase when the
characteristic measurement energy approaches the Planck energy.  In that
domain, the measurement errors of any observable are comparable to the
measurement values.

Limited measurement precision implies that at Planck energy it is impossible to
speak about points, instants, events\se, or dimensionality.  Limited precision
implies that no observable can be described by real numbers.  Limited
measurement precision also implies that at Planck length it is impossible to
distinguish positive and negative time values: particle and anti-particles are
thus not clearly distinguished at Planck scales.  A smallest length in nature
thus implies that there is no way to define exact boundaries of objects or
elementary particles.  But a boundary is what separates matter from vacuum.  In
short, a minimum measurement error means that at Planck scales, it is
impossible to distinguish objects from vacuum with complete
precision.\cite{me10}

So far, the conclusions drawn from the maximum force are essentially the same
that are drawn by modern research on unified theories.  The limit allows to
reach the same conceptual results found by string theory and the various
quantum gravity approaches.  To show the power of the maximum force limit, we
mention a few conclusions which go beyond them.

\subsubsection{Measurement precision and sets}

\noindent The impossibility to completely eliminate measurement errors has an
additional consequence.  In physics, it is assumed that nature is a \emph{set}
of components, all separable from each other.  This tacit assumption is
introduced in three main situations: it is assumed that matter consists of
(separable) particles, that space-time consists of (separable) events or
points, and that the set of states consists of (separable) initial conditions. 
So far, physics has built its complete description of nature on the concept of
\emph{set}.

A limited measurement precision implies that nature is \emph{not} a set of such
separable elements.\cite{me10,me11bis} A limited measurement precision implies
that distinction of physical entities is possible only approximately.  The
approximate distinction is only possible at energies much lower than the Planck
energy.  As humans we live at such smaller energies; thus we can safely make
the approximation.  Indeed, the approximation is excellent; we do not notice
any error when performing it.  But the discussion of the situation at Planck
energies shows that a perfect separation is impossible in principle.  In
particular, at the cosmic horizon, at the big bang, and at Planck scales any
precise distinction between two events or between two particles becomes
impossible.

Another way to reach this result is the following.  Separation of two entities
requires \emph{different measurement results}, such as different positions,
masses, sizes, etc.  Whatever observable is chosen, at Planck energy the
distinction becomes impossible, due to exploding measurements errors.  Only at
everyday energies is a distinction approximately possible.  Any distinction
among two physical systems, such as between a toothpick and a mountain, is thus
possible only \emph{approximately}; at Planck scales, a boundary cannot be
drawn.\cite{me10}

A third argument is the following.  In order to \emph{count} any entities
whatsoever -- a set of particles, a discrete set of points, or any other
discrete set of physical observables -- the entities have to be separable.  The
inevitable measurement errors make this impossible.  At Planck energy, it is
impossible to count physical objects with precision.\cite{me10} 

In short, at Planck energies a perfect separation is impossible in principle. 
We cannot distinguish observations at Planck energies.  In other words,
\emph{at Planck scale it is impossible to split nature into separate entities.}
There are no mathematical elements of any kind in nature.  Elements of sets
cannot be defined.  As a result, in nature, neither discrete nor continuous
sets can be constructed.\cite{me11bis}

Since sets and elements are only approximations, the concept of `set', which
assumes separable elements, is already too specialized to describe nature. 
Nature does not contain sets or elements.  The result implies that nature
cannot be described at Planck scales -- i.e., with full precision -- if any of
the concepts used for its description presupposes sets.  However, all concepts
used in the past twenty-five centuries to describe nature -- space, time, phase
space, Hilbert space, Fock space, particle space, loop space\se, or moduli
space -- are based on sets.  They all must be abandoned at Planck energy.  In
fact, \emph{no approach of theoretical physics so far, not even string theory
or the various quantum gravity approaches, satisfies the requirement to abandon
sets.} Nature has no parts.  \emph{Nature must be described by a mathematical
concept which does not contain any set.} This requirement can be used to guide
future searches for the unification of relativity and quantum theory.

\subsubsection{Hilbert's sixth problem}

\noindent In the year 1900, {David Hilbert} gave a well-known lecture in which
he listed\index{Hilbert's sixth problem} 
twenty-three of the great challenges facing mathematics in the
twentieth century.\cite{hiprlbis} Most problems provided challenges to many
mathematicians for decades afterwards.  Of the still unsolved ones, Hilbert's
sixth problem\index{Hilbert's problems}\index{Hilbert's sixth problem}
challenges mathematicians and physicists to find an \emph{axiomatic} treatment
of physics.  The challenge has stayed in the heads of many physicists since
that time.

Since nature does not contain sets, we can deduce that such an axiomatic
description of nature does not exist.\cite{me11bis} The reasoning is simple; we
only have to look at the axiomatic systems found in mathematics.  Axiomatic
systems define mathematical structures.  These structures come into three main
types: algebraic systems, order systems, or topological systems.  Most
mathematical structures -- such as symmetry groups, vector spaces, manifolds,
or fields -- are combinations of the three.  But all mathematical structures
contain sets.  Mathematics does not provide axiomatic systems which do not
contain sets.  The underlying reason is that every mathematical concept
contains at least one set. 

Also all physical concepts used in physics so far contain sets.  For humans, it
is already difficult to simply \emph{think} without first defining a set of
possibilities.  However, nature is different; nature does not contain sets. 
Therefore, \emph{an axiomatic formulation of physics is impossible.} Of course,
this conclusion does not rule out unification in itself; however, it does rule
out an axiomatic version of it.  The result surprises, as separate axiomatic
treatments of quantum theory or general relativity (see above) are possible. 
Indeed, only their unification does not allow one.  This is one of the reasons
for the difficulties of unified descriptions of nature.

\subsubsection{Outlook}

Physics can be summarized in a few limit statements.  They imply that in nature
every physical observable is limited by a value near the Planck value.  The
speed limit is equivalent to special relativity, the force limit to general
relativity, and the action limit to quantum theory.  Even though this summary
could have been made (or at least conjectured) by Planck, Einstein or the
fathers of quantum theory, it is much younger.  The numerical factors for
all limit values are new.  The limits provoke interesting Gedanken experiments;
none of them leads to violations of the limits.  On the other hand, the force
limit is not yet within direct experimental reach.

The existence of limit values to all observables implies that the description
of space-time with a continuous manifold is not correct at Planck scales; it is
only an approximation.  For the same reason, is predicted that elementary
particles are not point-like.  Nature's limits also imply the
non-distinguishability of matter and vacuum.  As a result, the structure of
particles and of space-time remains to be clarified.  So far, we can conclude
that nature can be described by sets only approximately.  The limit statements
show that Hilbert's sixth problem cannot be solved and that unification
requires fresh approaches, taking unbeaten paths into unexplored territory.

The only way to avoid sets in the description of nature seems to be to describe
empty space-time, radiation and matter as made of the same underlying entity. 
The inclusion of space-time dualities and of interaction dualities also seems
to be a necessary step.  Indeed, both string theory and modern quantum gravity
attempt this, though possibly not yet with the full radicalism necessary. 
The challenge is still open.


\end{document}